\documentclass[final,5p,twocolumn]{elsarticle}

\biboptions{sort&compress}
\usepackage[utf8]{inputenc}

\usepackage{bm}
\usepackage{amsmath,amssymb}
\usepackage{newtxtext}
\usepackage{newtxmath}

\usepackage{graphicx}
\usepackage{color}
\usepackage{float}
\usepackage{textcomp}
\usepackage{esint}
\usepackage{multirow}
\usepackage{longtable}
\usepackage{array}
\usepackage{booktabs}
\usepackage{siunitx}
\usepackage{tabularx}
\usepackage{slashed}
\usepackage{subfigure}
\usepackage{verbatim}
\usepackage[hidelinks]{hyperref}

\newcolumntype{C}{>{\centering\arraybackslash}X}

\hypersetup{
  pdftitle={SU4Interaction},
  pdfauthor={BNLU},
  unicode=true,
  bookmarksnumbered=false,
  bookmarksopen=false,
  breaklinks=false,
  pdfborder={0 0 1},
  colorlinks=true,
  allcolors=blue
}

\journal{Physics Letters B}

\begin{document}

\begin{frontmatter}

\title{Extracting Resonance Width from Lattice Quantum Monte Carlo Simulations \\
Using Analytical Continuation Method}


\author[inst1]{Zhong-Wang Niu}

\author[inst2]{Shi-Sheng Zhang\corref{cor1}}

\cortext[cor1]{Corresponding authors.}
\ead{zss76@buaa.edu.cn}

\author[inst1]{Bing-Nan Lu\corref{cor1}}
\ead{bnlv@gscaep.ac.cn}
\affiliation[inst1]{
  organization={Graduate School of China Academy of Engineering Physics},
  city={Beijing},
  postcode={100193},
  country={China}
}

\affiliation[inst2]{
  organization={School of Physics, Beihang University},
  city={Beijing},
  postcode={100191},
  country={China}
}






\begin{abstract}
Nuclear lattice effective field theory (NLEFT) provides an efficient \textit{ab initio} framework for computing low-lying states via imaginary time projection. However, the extraction of unstable resonances, especially those with broad widths, remains a significant challenge. Traditional techniques such as the complex scaling method are often limited by sign problems or inherent statistical uncertainties. In this work, we present the first direct extraction of a nuclear resonance width within NLEFT, by combining a high-precision, sign-problem-free nuclear interaction with the analytical continuation in the coupling constant (ACCC) approach. To address numerical instabilities in the ACCC framework, we implement a robust Pad\'e solver based on singular value decomposition (SVD), incorporating ridge regularization and pole-safety criteria to ensure reliable extrapolation to the resonance pole. We detail the methodology and apply it to the unbound ground state of $^5\mathrm{He}$ ($J^\pi = 3/2^-$). Our calculation yields a resonance energy $E = 0.80(10)$ MeV and a width $\Gamma = 1.05(9)$ MeV, in agreement with recent experimental results ($E_{\mathrm{exp}} = 0.798$ MeV, $\Gamma_{\mathrm{exp}} = 0.648$ MeV). This work establishes a practical and precise strategy for studying resonances within the \textit{ab initio} lattice framework, paving the way for investigations of many-body resonances in exotic nuclei near the drip lines.

\end{abstract}

\end{frontmatter}

\section{Introduction}

Nuclear Lattice Effective Field Theory (NLEFT) is an efficient \textit{ab initio} method utilizing the auxiliary-field quantum Monte Carlo (QMC) approach. Significant progress has been made with NLEFT in recent years, enabling studies of nuclear ground and excited states~\cite{EPJA31-105,PRL104-142501, EPJA45-335, PLB732-110, PRL112-102501, PLB869-139839,PRL134-162503}, clustering phenomena~\cite{PRL106-192501, PRL109-252501,PRL110-112502,PRL119-222505}, hypernuclei~\cite{EPJA60-215,PRD111-036002}, scattering~\cite{Nature528-111,PRC86-034003,JPGPP52-125102},  thermodynamics~\cite{PRL115-185301},  and electro-weak reactions~\cite{PLB859-139086,PRC112-025502}. Within NLEFT, the nuclear Hamiltonian is discretized on a spatial cubic lattice and solved using imaginary-time projection to obtain low-lying eigenstates.
Most NLEFT studies to date have focused on the bound states or treated very narrow resonances as the bound states approximately. For example, the energy of the second $0^+$ state in $^{12}$C (Hoyle state), which plays a crucial role in stellar nucleosynthesis~\cite{Hoyle1954}, lies just above the three-$\alpha$ threshold. It is a resonance with an extremely narrow width of a few electron-volts against decay into three $\alpha$-particles. In such cases, direct NLEFT simulations can reliably extract its energy~\cite{,PRL106-192501, PRL109-252501,PRL110-112502}. However, these calculations provide no information on the resonant width and cannot be directly applied to broad resonances. So far, precisely describing general resonance properties remains particularly challenging within NLEFT.

Generally, the single-particle resonances in the potential model correspond to the poles of $S$-matrix. A variety of theoretical approaches have been established to extract resonance parameters, $\it{i.e.}$ resonance energy $E$ and width $\Gamma$.
These include the $R$-matrix and $K$-matrix formalisms~\cite{PR72-29,PRL59-763,PRC44-2530,Taylor2012,PRL109-072501,PRC88-024323,PLB850-138511} based on fitting to scattering or reaction observables,
the Green's function techniques~\cite{SJNP45-783,PRC90-054321,CPC44-084105},
the complex scaling method (CSM)~\cite{PRC34-95,PRC37-383,PRL79-2217}, 
the complex momentum representation method~\cite{PRC73-034321,PRL117-062502,PLB801-135174,NST2023Xu}, 
the real stabilization method~\cite{PRA1-1109,PRC77-014312} and the phase-shift method~\cite{PRC2003Sandulescu} suitable for very narrow resonance, as well as the practical analytical continuation in the coupling constant (ACCC) method especially suitable for broad resonances~\cite{PRC70-034308,EPJA48-40,PLB730-30,PRC92-024324,CCP21-1154,CPC49-014108}, etc.

Resonance and continuum properties in few-body nuclei have increasingly been studied with \textit{ab initio} methods based on realistic or chiral nuclear interactions, including the no-core shell model with continuum, the no-core Gamow shell model, and SS-HORSE analyses~\cite{PRC87-034326,PRC88-044318,PRC106-064320}. Related progress has also been achieved in Gamow-based many-body approaches~\cite{PRC99-061302,PRC104-024319}.
Complementary \textit{ab initio} studies based on the Faddeev--Yakubovsky (FY) formalism have addressed five-body $n$--$\alpha$ scattering relevant to the resonant channels of $^{5}\mathrm{He}$, as well as broad $^{5}\mathrm{H}$ resonance poles in the complex energy plane~\cite{FrontPhys7-251,PRC97-044002,PLB791-335}.
An $\textit{ab initio}$ approach to open quantum systems has been implemented for light nuclei by using a Gamow–Hartree–Fock basis and realistic interactions and employing coupled-cluster theory to solve the quantum many-body problem for the helium chain~\cite{PLB656-169}.
Other $\textit{ab initio}$ methods like the Green's function Monte Carlo (GFMC)~\cite{Carlson2005} or the no-core shell model~\cite{PRC-Caurier2006} have been used to calculate the structure of helium isotopes, which allows nonperturbative calculations with realistic potentials but often shows larger deviations for resonance observables ~\cite{wiringa1995}. 

In lattice quantum chromodynamics, L\"uscher-type finite-volume methods connect the discrete finite-volume spectrum to scattering amplitudes and resonance parameters, providing a first-principles framework for studying elastic and coupled-channel hadronic resonances~\cite{PRD83-094505,PRD87-034505,PRD91-054008}.
Resonances appear as plateaus, with slopes indicating widths. This finite-volume approach was recently applied in NLEFT using Wigner-SU(4) and N$^3$LO chiral forces to study the plausible tetraneutron resonance~\cite{arXiv:2601.01801}.
An efficient alternative involves computing scattering-amplitude poles in the complex-energy plane, where pole positions directly yield resonance energies (real parts) and widths (imaginary parts). 
These methods provide more direct resonance signatures and better resolve complex structures like overlapping poles or broad resonances. 
However, they require analytical continuation from real to imaginary momentum, which is a typically ill-posed problem introducing significant numerical uncertainties.
Precise calculations thus demand both high-accuracy numerics and efficient analytical-continuation algorithms.

QMC methods like NLEFT inherently exhibit statistical uncertainties that grow exponentially with sign problems~\cite{PRL94-170201}. 
Most NLEFT calculations treat the leading-order interaction non-perturbatively with minimal sign problem and handle residual interactions perturbatively.
While this reduces statistical errors, the perturbative truncation error remains uncontrolled, preventing reliable analytical continuation.
Recently, Ref.~\cite{PRL135-222504} introduced LAT-OPT1, a high-precision, sign-problem-free lattice interaction. 
It provides excellent predictions for ground-state properties from helium to tin isotopes and accurately reproduces shell structures through a sign-problem-free spin-orbit term. 
For odd-even nuclei like $^5$He, a mild sign problem persists.
Nevertheless, we expect that LAT-OPT1 still enables fully non-perturbative calculations meeting the precision demands of complex-energy methods for resonances.

$^5$He is an unbound system in the $\alpha{+}n$ channel dominated by broad $P$-wave resonances. The spin--orbit interaction generates $J^\pi=3/2^-$ and $1/2^-$ states governing low-energy $n$--$\alpha$ scattering~\cite{LandoltBornstein2012,NPA708-3}. As a short-lived compound nucleus, it induces near-threshold enhancement in $D{+}T\to \alpha{+}n$ cross sections, which is critical for fusion ignition and polarized-fuel optimization with similar effects in channels like $T{+}T$ fusion~\cite{PRL133-055102,NC10-351,FST80,arXiv2305-00647,PRC109-054620}. The absence of bound $A=5$ nuclei constrains nucleosynthesis, delaying element production beyond helium until multi-particle processes activate~\cite{arXiv2305-00647,IJMPE28-1930004}. 
Accurate description of $^5$He is of great physical importance and requires consistent treatment of spin-orbit coupling, three-body forces, continuum effects and $\alpha{+}n$ clustering, imposing rigorous constraints on nuclear forces~\cite{PRC102-024616,PR890-1,JPGPP52-125102} and many-body methods. 
Ref.~\cite{PRC102-024616} obtained $n$--$\alpha$ phase shifts and resonance parameters via no-core shell model (NCSM), where $^5$He data was applied to constrain the three-nucleon forces. 
Ref.~\cite{JPGPP52-125102} computed phase shifts using lattice Monte Carlo, though statistical uncertainties precluded precise width extraction.

In this work, we present the first calculation of the $^5$He ground-state energy and width by combining the LAT-OPT1 interaction with the analytic continuation of the coupling constant method. 
To ensure numerical stability, we implement ACCC via a singular value decomposition (SVD)-based Pad\'e solver, which regularizes the underlying linear system.
We further provide a data-driven uncertainty estimate for the resonance parameters. 
This work establishes a practical strategy for resonance extraction within the lattice auxiliary-field Monte Carlo framework.

\section{Theoretical framework}

We use natural units with $\hbar=c=1$ throughout.
The Hamiltonian in LAT-OPT1 is defined on a $L^3$ cubic lattice with integer coordinates $\bm{n}=(n_x,n_y,n_z)$,
\begin{eqnarray}
    H = \sum_{\bm{n}} \left[ -\frac{\Psi^\dagger \nabla^2 \Psi}{2M} + :\frac{C_2}{2} \overline{\rho}^2 + \frac{C_3}{6} \overline{\rho}^3 + \frac{C_s}{2} \overline{\rho}\ \overline{\rho}_{s}: \right],
    \label{eq:Hamiltonian}
\end{eqnarray}
where the sum runs over all lattice sites and $:\cdots:$ denotes normal ordering. 
Here $\Psi(\bm{n})$ and $\Psi^\dagger(\bm{n})$ are nucleon annihilation and creation operators, $M$ is the nucleon mass, and $\nabla^2$ is the lattice Laplacian. 
The coupling constants $C_2$, $C_3$, and $C_s$ multiply the smeared local densities $\overline{\rho}$ and $\overline{\rho}_s$ (overlines denote the smearing used in LAT-OPT1). 
The detailed smearing prescription and the AFQMC sampling procedure used to obtain the projected wave functions and observables are given in Ref.~\cite{PRL135-222504}.

Resonant states correspond to poles of the scattering amplitude on the complex-energy plane~\cite{Taylor2012}. 
Such poles are not obtained directly from solving for real eigenenergies, and they are also not accessed by solving the scattering equation only on the real-energy axis. 
Analytic continuation provides a practical route to determine the pole position from information in a region where the solution is well defined. 
As a two-body example, we consider the stationary radial Schr\"odinger equation
\begin{equation}
\left[\frac{\mathrm{d}^2}{\mathrm{d}r^2} - \frac{l(l+1)}{r^2} -2M V(r) + p^2 \right]\varphi_l(p,r)=0,
\label{eq:Sch}
\end{equation}

Introducing a coupling constant $\lambda$, the Hamiltonian of the scattering system can be written as $H = H_0 + \lambda V$. The branch point $\lambda_{0}$ denotes the critical value separating the bound states and the continuum states.

The behavior of the momentum $\it{p}$ near $\lambda = \lambda_0$ can be approximated by the expression as follows,
\begin{align}
p(\lambda) \sim
\begin{cases}
 -i(\lambda - \lambda_{0}), & l = 0; \\[6pt]
 \pm i\sqrt{\lambda - \lambda_{0}}, & l > 0.
\end{cases}
\label{eq:momentum_beheavier}
\end{align}
When the attractive potential $\lambda V(r)$ is deeper than $\lambda_0 V(r)$, there exists a zero on the positive imaginary axis of the complex momentum plane, corresponding to a bound state.
As the attraction is reduced so that $\lambda$ approaches $\lambda_0$, this zero gradually moves toward the origin. Upon further decreasing the attraction, the zero leaves the origin and moves into the lower half-plane of the complex momentum plane. When $\lambda$ is reduced to $\lambda < \lambda_0$, this zero corresponds to a resonance.

Since the resonance corresponds to the pole in the complex momentum plane, the Pad\'e approximant of the second kind (PAII)~\cite{Baker1996} 
with the form of a rational function is adopted, which provides a practical tool for analytical continuation of the momentum $p$ into the complex plane.
The PAII $f^{[N,M]}(z)$ of a function $f(z)$ is defined as
\begin{equation}
f^{[N,M]}(z) = \frac{P_N(z)}{Q_M(z)} = f(z) + O\!\left(z^{N+M+1}\right),
\end{equation}
where $P_N(z)$ and $Q_M(z)$ are polynomials of degree $N$ and $M$, respectively, whose coefficients are uniquely determined by the first $N+M+1$ coefficients of $f(z)$.

Given the complex momentum $k=k_r+i k_i$ at the physical point $\lambda=1$, the resonance energy $E$ and width $\Gamma$ are expressed in the nonrelativistic limit as 
\begin{equation}
E = \frac{k_r^2-k_i^2}{2\mu}, \qquad \Gamma = -\frac{2k_rk_i}{\mu}. 
\end{equation}
Within this framework, the ACCC method uses high-precision bound-state information as input to infer resonance properties through analytic continuation. 
In the present work, this strategy becomes feasible because the sign-problem-free LAT-OPT1 interaction enables AFQMC calculations to provide bound-state energies with sufficiently high accuracy, which is essential for a stable continuation to the resonance pole.

\section{Numerical Implementation}\label{sec:pade}

The procedure can be roughly divided into the following steps: First, we determine the branch point $\lambda_0$ by $E(\lambda_0) = 0$.
Next, for $\lambda > \lambda_0$, we choose an appropriate interval in $\lambda$ and compute the corresponding energy levels for bound states. Finally, we perform an analytic continuation of the momentum using a PAII to obtain the position of the resonance pole in complex momentum plane for $\lambda=1$. In such a way, we can extract the resonance parameters and analyze the associated uncertainties.

\subsection{Determination of the Branch Point}
The branch point $\lambda_0 = 1.1331$ can be achieved by solving $p(\lambda_0)=0$ with the bisection method. 
According to the behavior of $p$ near $\lambda = \lambda_0$ in Eq.~(\ref{eq:momentum_beheavier}),
we perform the change of variables $z = \sqrt{\lambda - \lambda_{0}}$ for the case $l>0$.

\subsection{Acquisition of Bound-State Energy Data}
In the interval $z\in[0.1,0.6]$, we choose a set of uniformly spaced points and compute the ground-state energies $E_{^{4}\mathrm{He}}(z)$ and $E_{^{5}\mathrm{He}}(z)$ with LAT-OPT1 on a lattice of size $L=11$. In the bound-state region, we then define the relative energy with respect to the $n+\alpha$ threshold by
\begin{equation}
E_b(z_i)\equiv E_{^{5}\mathrm{He}}(z_i)-E_{^{4}\mathrm{He}}(z_i)
= -\frac{\kappa(z_i)^2}{2\mu},
\end{equation}
where $\mu$ is the reduced mass of the $n+\alpha$ system. This yields the input data set $\{z_i,\kappa(z_i)\}$ for the subsequent analytic continuation.
The resonance pole is located at $z_{\star}$($\lambda=1$). Full data set is provided in Supplemental Material~\cite{SM}.

\subsection{Analytic Continuation}

Given the bound-state data set, we use a PAII to approximate the function $\kappa(z)$ and then analytically continue it to the physical point $z_{\star}$. In the bound-state region, the input values $\kappa(z_i)$ are real and positive. The ansatz is
\begin{equation}
\kappa(z) \approx \frac{P_N(z)}{Q_M(z)}
=\frac{\sum_{p=0}^{N} c_p z^p}{\sum_{q=0}^{M} d_q z^q},
\end{equation}
with
$c_0=0$ and $d_0=1$.
The explicit construction of the linear system for the Pad\'e coefficients and the corresponding least-squares solution are given in Supplemental Material~\cite{SM}.

Fig.~\ref{fig:k_analytic_structure} shows the trajectory of the poles in the complex momentum plane as
$z$ or $\lambda$ varies, moving from the imaginary axis [$p(z_i)$ for the bound states] into the fourth quadrant [$p(z_{\star})$ for the resonant state].
This behavior is fully consistent with the predictions of scattering theory.

\begin{figure}[htbp]
    \centering
    \includegraphics[width=\columnwidth]{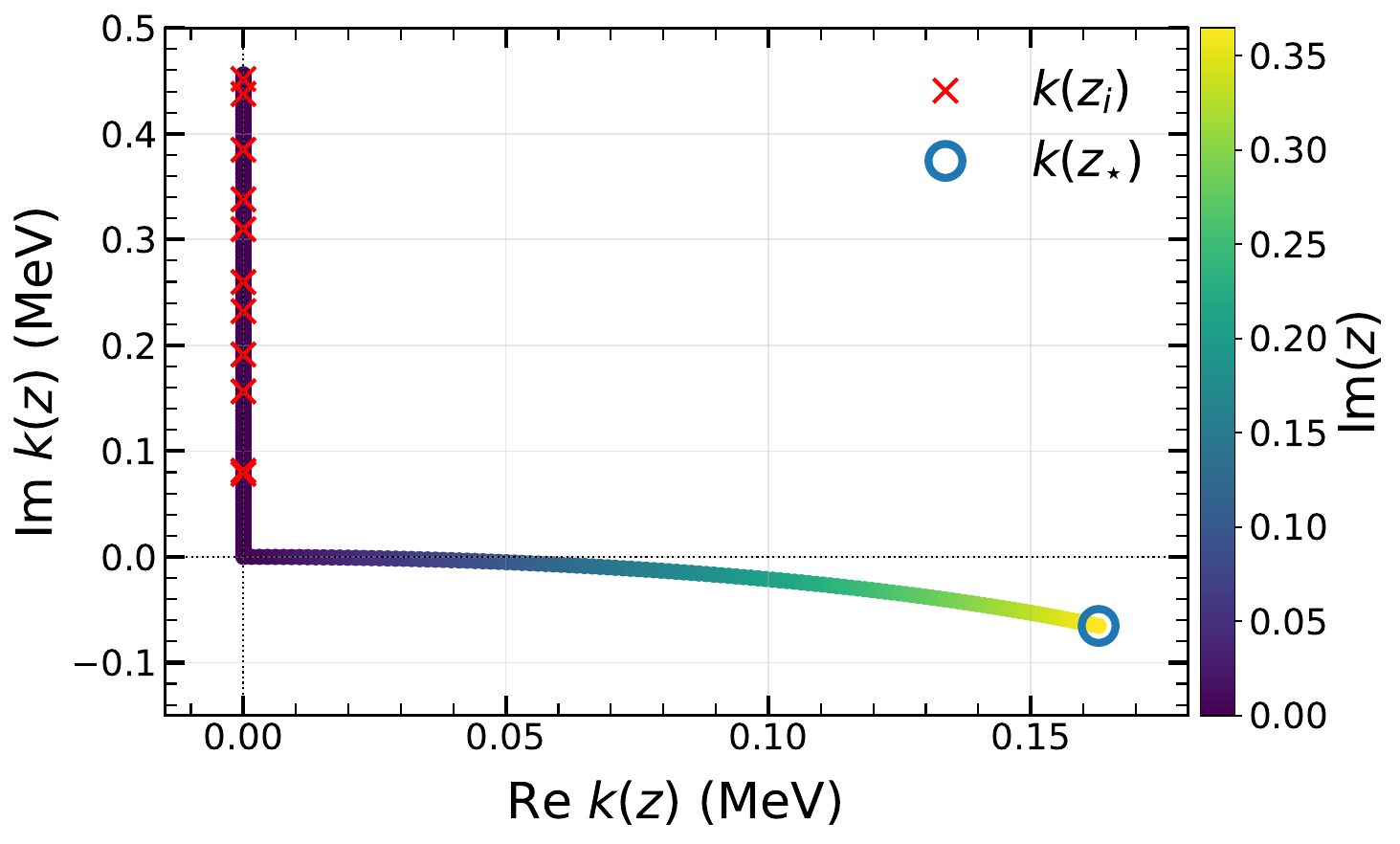}
    \caption{
        Analytic continuation of the complex momentum $k(z)=i\kappa(z)$ in the complex momentum plane. As $\mathrm{Im}(z)$ varies, $k(z)$ traces the colored trajectory (color bar: $\mathrm{Im}(z)$), starting from the bound-state input points $k(z_i)$ on the positive imaginary axis (red crosses), passing through the threshold point $k=0$, and ending at the resonance pole $k(z_\star)$ (open blue circle) in the fourth quadrant.
    }
    \label{fig:k_analytic_structure}
\end{figure}

\subsection{Treatment of ill-conditioning}

The Pad\'e fitting problem is intrinsically ill-conditioned due to the Vandermonde-type structure of the design matrix.
The direct consequence of such ill-conditioning is extremely sensitive to data and noise: tiny perturbations in the input can cause large fluctuations in the coefficients, which then manifest in the rational function as pole drift, near zero--pole pairs, and nonphysical oscillations in extrapolation. In analytic continuation problems, this severely contaminates the inference of singularities (including resonance poles) and the estimation of their uncertainties.

To stabilize the Pad\'e fitting procedure, we analyze the linear system using SVD~\cite{Golub1970,Gonnet2013} and quantify its numerical stability by the spectral condition number. 
We then apply a two-step regularization strategy. 
First, column equilibration is employed to remove trivial scale disparities among different polynomial orders. 
Second, ridge (Tikhonov) regularization\cite{Tikhonov1963,Hoerl1970} with a regularization parameter $\lambda_T$ is introduced to suppress near-linear dependencies and damp noise amplification in poorly constrained directions.
The detailed construction of the regularized linear system is provided in Supplemental Material~\cite{SM}.

Figure~\ref{fig:pade55_spurious_poles} illustrates the Pad\'e $(5,5)$ approximation as a representative example, showing how numerical ill-conditioning is alleviated by the proposed stabilization procedure. 
After column equilibration, an appropriate choice of the ridge parameter $\lambda_T$ effectively suppresses the emergence of nonphysical spurious poles and yields a stable fit.

\begin{figure}[htbp]
    \centering
    \includegraphics[width=\columnwidth]{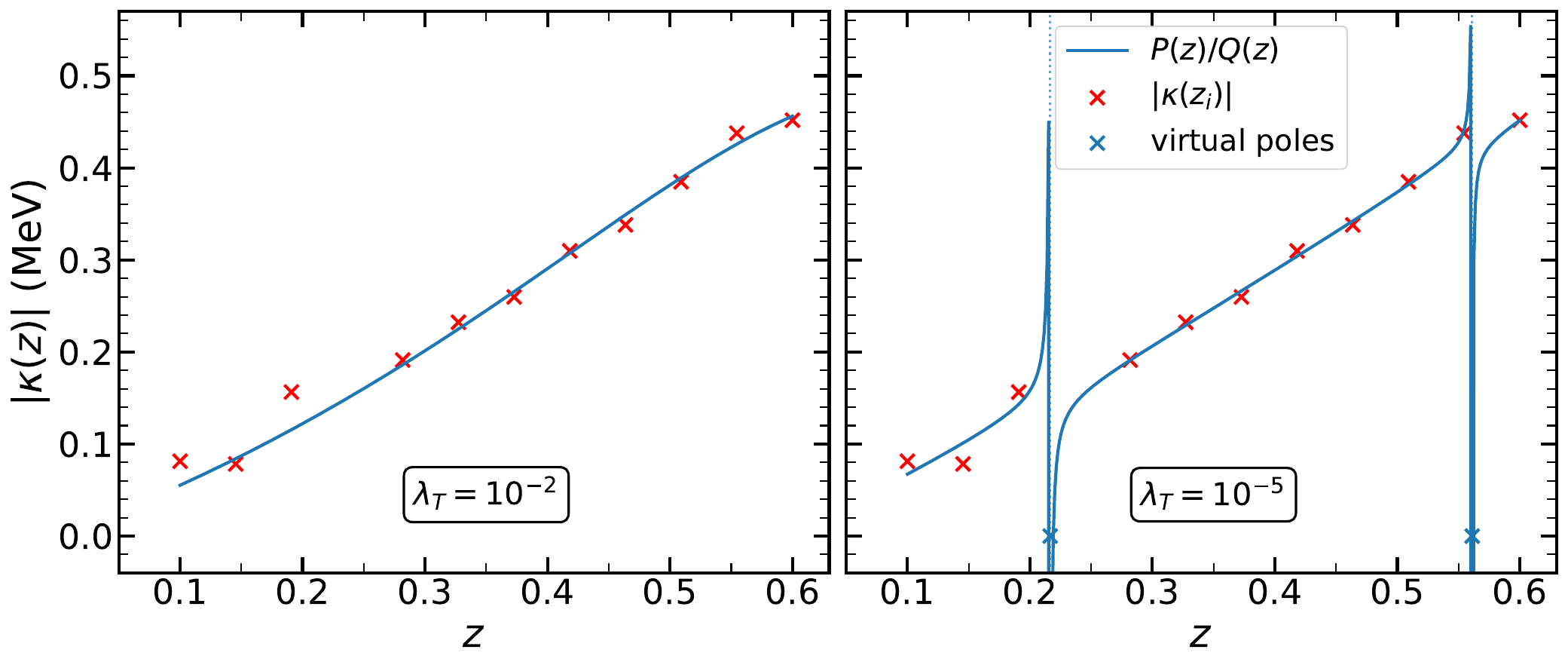}
    \caption{
        Comparison of Pad\'e $(5,5)$ approximants obtained with two different ridge
        regularization strengths.
        The blue curves show the fit function values $P(z)/Q(z)$,
        while red crosses denote the input data points $\kappa(z)$.
        Left panel: for $\lambda_T = 10^{-2}$, the Pad\'e approximant remains smooth
        and free of real-axis poles within the fitting interval.
        Right panel: for a much smaller regularization, $\lambda_T = 10^{-5}$,
        several spurious poles (marked by blue crosses) enter the
        interpolation domain, producing sharp unphysical spikes in $P(z)/Q(z)$
        .
    }
    \label{fig:pade55_spurious_poles}
\end{figure}

Numerical instability in Pad\'e-based analytic continuation typically manifests first in the pole structure.
We therefore monitor the distribution of poles and identify spurious poles as those unrelated to physical singularities or expected branch points, but generated by noise amplification in ill-conditioned fits.

The poles are given by the roots of the Pad\'e denominator $Q_M(z)$.
Let us define $z_{\max}=\max_i |z_i|$ and $S=\max(|z_\star|,z_{\max})$ as a characteristic scale in the complex-$z$ plane.

To ensure stability, we impose two pole-safety criteria.
First, fits are rejected if a pole appears on (or arbitrarily close to) the real axis within the training interval.
Second, fits are discarded if any pole approaches the target point too closely, quantified by the normalized distance
\begin{equation}
d_{\min}^{(\mathrm{norm})}=\frac{\min_j |z^{(\mathrm{pole})}_j-z_\star|}{S},
\end{equation}
and requiring $d_{\min}^{(\mathrm{norm})}>\tau_{\mathrm{pole}}$, where $\tau_{\mathrm{pole}}$ is a small positive real parameter.
Only fits satisfying both criteria are retained.

\subsection{Assessment of Fit Quality}

To assess the quality and stability of the Pad\'e fits, we employ the root-mean-square error (RMS) and leave-one-out cross validation (LOOCV).
Given the fitted Pad\'e model, the predicted values at the training points are denoted by $\widehat{y}_i$.
The RMS is defined as
\begin{equation}
\mathrm{RMS}=\sqrt{\frac{1}{K}\sum_{i=1}^K\bigl(\widehat{y}_i-y_i\bigr)^2},
\end{equation}
which quantifies the overall agreement with the training data.
The LOOCV is defined as
\begin{equation}
\mathrm{LOOCV}
=\sqrt{\frac{1}{K}\sum_{i=1}^K\Bigl(\widehat{y}^{(-i)}(z_i)-y_i\Bigr)^2},
\end{equation}
where $\widehat{y}^{(-i)}(z_i)$ denotes the prediction obtained by refitting the model with the $i$-th data point removed.
While RMS measures the fit quality on the training set, LOOCV probes the stability of the fit against data removal and thus provides a diagnostic of overfitting and numerical instability.
In practice, a very small RMS accompanied by a much larger LOOCV indicates overfitting or noise amplification, often correlated with unstable pole behavior, whereas comparable RMS and LOOCV values signal a well-balanced fit.
Conversely, large values of both metrics suggest an overly rigid model with insufficient flexibility.

Ridge regularization suppresses noise amplification in poorly constrained directions, but the regularization strength $\lambda_T$ must be determined in a data-driven manner.
We scan $\lambda_T$ over a logarithmic range and retain only those fits that satisfy the pole-safety criteria.
Within this pole-safe region, $\lambda_T$ is chosen at the minimum of the RMS and LOOCV curves, corresponding to a balance between data fidelity and numerical stability.

\begin{figure}[htbp]
    \centering
    \includegraphics[width=\columnwidth]{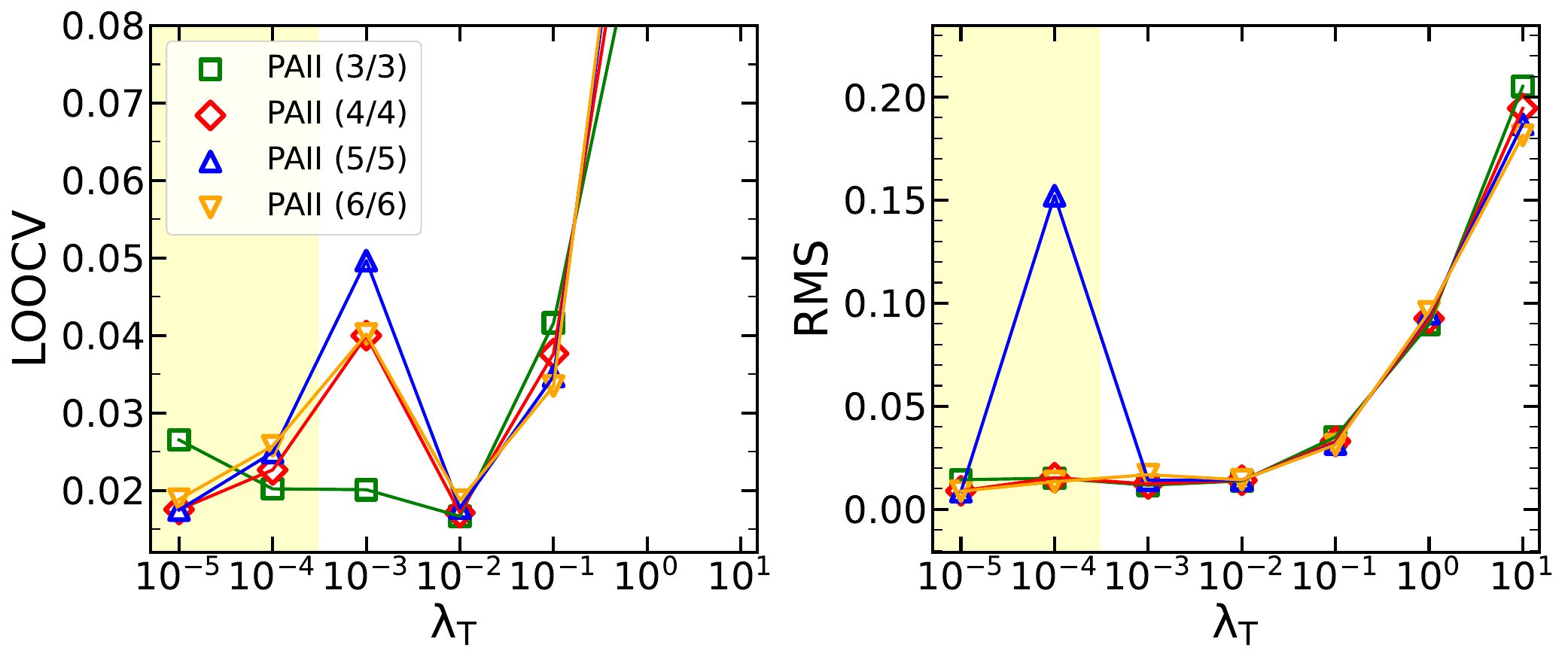}
    \caption{
        Dependence of LOOCV (left panel)
        and RMS (right panel) on the ridge
        regularization parameter $\lambda_T$ for
        PAII approximants of different orders.
        Squares, diamonds, upward triangles, and downward triangles correspond
        to the PAII $(3/3)$, $(4/4)$, $(5/5)$, and $(6/6)$ approximants,
        respectively.
        The light yellow shaded region marks the small-regularization regime in which spurious poles emerge.
    }
    \label{fig:cv_ridge_pade}
\end{figure}

Figure~\ref{fig:cv_ridge_pade} illustrates this procedure for representative Pad\'e orders.
A clear minimum of both metrics is observed around $\lambda_T\sim10^{-2}$, which we identify as the ``safe point in the valley.''
Smaller values of $\lambda_T$ lead to noise amplification and unstable pole structures, whereas larger values result in overly rigid fits.

\subsection{Jackknife stability estimate for $(E,\Gamma)$}\label{sec:loo-uncertainty}

We employ jackknife analysis~\cite{Quenouille1949,Quenouille1956,Tukey1958} to estimate the uncertainty of the extracted resonance energy and width.
For a fixed Pad\'e order $(N,M)$ and regularization strength $\lambda_T$, we construct $K$ leave-one-out data sets by removing one sampling node at a time,
\begin{equation}
\mathcal{D}^{(-i)} \equiv \{z_j,\kappa(z_j)\}_{j\neq i}, \qquad i=1,\dots,K,
\end{equation}
and repeat the same regularized Pad\'e continuation (with identical pole-safety criteria) to obtain $(E^{(-i)},\Gamma^{(-i)})$ at the physical point $z_\star$.

The jackknife variance is then evaluated as
\begin{equation}
\sigma_\theta^2
= \frac{K-1}{K}\sum_{i=1}^{K}\left(\theta^{(-i)}-\bar{\theta}\right)^2,
\qquad
\bar{\theta}=\frac{1}{K}\sum_{i=1}^{K}\theta^{(-i)},
\end{equation}
with $\theta\in\{E,\Gamma\}$.
The resulting $(\bar{E}\pm\sigma_E,\;\bar{\Gamma}\pm\sigma_\Gamma)$ provides a data-driven estimate of the continuation stability associated with the discrete sampling of the bound-state region.

\section{Results and discussion}\label{sec:results}

Table~\ref{tab:pade-c0} summarizes the resonance parameters extracted from Pad\'e approximants of orders $(3,3)$--$(6,6)$.
The fit-quality diagnostics (RMS and LOOCV) remain small and close to each other across all Pad\'e orders,
indicating stable fits and good extrapolation behavior.
In principle, different Pad\'e orders do not form a convergent sequence, so we cannot claim that any single result between $(3,3)$ and $(6,6)$ is more reliable.
Nevertheless, the extracted resonance parameters show only modest variations among these orders.
We therefore quote a representative estimate by combining the results in Table~\ref{tab:pade-c0} with an inverse-variance weighted average over Pad\'e orders,
and include the residual Pad\'e-order scatter as an additional systematic uncertainty:
$E = \num{0.80(10)}~\mathrm{MeV}$, $\Gamma = \num{1.05(9)}~\mathrm{MeV}.$
Compared with the experimental reference values for the $^5$He $J^\pi=3/2^-$ resonance,
$E_{\mathrm{exp}} = 0.798~\mathrm{MeV}$ and $\Gamma_{\mathrm{exp}}= 0.648~\mathrm{MeV}$,
our results are compatible within the theoretical uncertainty currently allowed by LAT-OPT1 and the continuation uncertainty estimated by jackknife.

\begin{table}[htbp]
\centering
\small
\setlength{\tabcolsep}{8pt}
\caption{
Resonance parameters extracted from Pad\'e models of orders $(3,3)$ through $(6,6)$.
The resonance energy $E$ and width $\Gamma$ are reported together with their
uncertainties, shown in parentheses, which are estimated using jackknife
analysis.
Also listed are the LOOCV and RMS values used to assess the fit quality.
All reported results satisfy the pole-safety criteria.
}

\label{tab:pade-c0}
\begin{tabular}{lcccc}
\toprule
\textbf{Order} & \textbf{$E$ (MeV)} & \textbf{$\Gamma$ (MeV)} & \textbf{LOOCV} & \textbf{RMS} \\
\midrule
$\text{(3,3)}$ & \num{0.80(21)} & \num{1.06(10)} & \num{0.0163} & \num{0.0122}  \\
$\text{(4,4)}$ & \num{0.80(21)} & \num{1.07(17)} & \num{0.0184} & \num{0.0121}  \\
$\text{(5,5)}$ & \num{0.80(19)} & \num{1.02(23)} & \num{0.0227} & \num{0.0121}  \\
$\text{(6,6)}$ & \num{0.80(18)} & \num{0.98(22)} & \num{0.0252} & \num{0.0121}  \\
\bottomrule
\end{tabular}
\end{table}

Based on the analysis, the total uncertainty mainly has two sources.
First, it is set by the theoretical accuracy of LAT-OPT1 together with the statistical uncertainty of the auxiliary-field Monte Carlo ground-state energies.
We also use a relatively large lattice size, $L=11$, so the finite-volume effect is smaller than the Monte Carlo statistical error and can be neglected at the present precision.
Second, additional uncertainty arises from the ACCC analytic continuation, including the Pad\'e fit error, the dependence on the chosen sampling window, the intrinsic sensitivity of continuation, and the residual model dependence across Pad\'e orders.
We quantify this continuation component with a jackknife analysis.

In this work, we propose a practical framework that combines LAT-OPT1, a high-precision and solvable lattice nuclear interaction, with the ACCC method to extract resonance parameters from bound-state data. Using the $^{5}$He $J^\pi=3/2^-$ resonance as a test case, we demonstrate the full workflow and obtain resonance parameters with controlled uncertainties. We also analyze in detail the spurious-pole problem caused by ill-conditioning in Pad\'e-based continuation, and we alleviate it using column equilibration, ridge regularization, and suitable pole-safety criteria. With these modifications, we obtain resonance parameters with relatively small uncertainties. Because analytic continuation needs to extend the pole from the positive imaginary axis in momentum space (the bound-state region) to the lower half-plane (the resonance region), high-precision Monte Carlo input for the bound-state data is essential. The per-mille-level statistical uncertainty enabled by the sign-problem-free nature of LAT-OPT1 is therefore crucial for accurate continuation.

\section{Outlook}\label{sec:outlook}
This framework can be applied to more complex resonance problems, such as excited resonances, many-body resonances, and larger unstable systems that are difficult to treat with traditional \textit{ab initio} methods. Since many nuclei are unstable, accurate first-principles resonance calculations would greatly expand the range of systems accessible to lattice auxiliary-field Monte Carlo. This provides a useful tool for future studies of exotic nuclei near the drip lines.

\section*{Acknowledgements}
Z.W.~Niu and Dr.~B.N.~Lu was supported by NSAF No. U2330401 and National Natural Science Foundation of China with Grant No. 12275259, 12547105 and the Science Challenge Project (No. TZ2025012). 
Dr.~S.S.~Zhang was supported by the National Natural Science Foundation of China (Grant Nos.~12575122).


\clearpage
\onecolumn
\setcounter{table}{0}
\renewcommand{\thetable}{S\arabic{table}}

\section*{Supplementary Material}

\setcounter{equation}{0}
\renewcommand{\theequation}{S\arabic{equation}}

This Supplemental Material provides technical details that support the main text. 
In particular, we present (i) the bound-state energy data set obtained with LAT-OPT1 and the corresponding construction of momenta used in the ACCC analysis, 
(ii) the explicit linear formulation employed to determine the Pad\'e coefficients from the bound-state data, and 
(iii) the SVD-based stabilization procedure, including column equilibration and ridge (Tikhonov) regularization.

\section*{Bound-State Energy Data}

We provide the bound-state energy data used in the ACCC analysis. 
For each value of the auxiliary coupling parameter $\lambda$, the ground-state energies of
$^{4}\mathrm{He}$ and $^{5}\mathrm{He}$ are computed using the LAT-OPT1 interaction on a lattice of size $L=11$. 
The calculations are performed independently for the two nuclei at the same value of $\lambda$.

From the resulting energies $E_{^{4}\mathrm{He}}(\lambda)$ and $E_{^{5}\mathrm{He}}(\lambda)$, we define the relative energy with respect to the $n+\alpha$ threshold as
\begin{equation}
E_b(\lambda)\equiv E_{^{5}\mathrm{He}}(\lambda)-E_{^{4}\mathrm{He}}(\lambda),
\end{equation}
which is negative in the bound-state region. The corresponding binding momentum $\kappa(\lambda)>0$ is then obtained from
\begin{equation}
E_b(\lambda)=-\frac{\kappa(\lambda)^2}{2\mu},
\end{equation}
where $\mu$ is the reduced mass of the $n+\alpha$ system, as defined in the main text.
The tabulated values of $\kappa$ are therefore derived quantities calculated from the energy differences, rather than direct outputs of the lattice calculation.

Table~\ref{tab:lambda-energy} lists the ground-state energies of $^{4}\mathrm{He}$ and $^{5}\mathrm{He}$, together with the corresponding values of $\kappa$, for all coupling parameters $\lambda$ used in the Pad\'e analysis.
These data constitute the input set $\{(\lambda_i,\kappa(\lambda_i))\}$ employed for the analytic continuation to the physical point $\lambda=1$.

\begin{table}[htbp]
\centering
\small
\setlength{\tabcolsep}{6pt}
\caption{
Bound-State Energy Data.
}
\label{tab:lambda-energy}
\begin{tabular}{c|ccccccccccc}
\toprule
$\lambda$ 
& 1.142 & 1.153 & 1.168 & 1.211 & 1.239 & 1.271 & 1.307 & 1.347 & 1.391 & 1.440 & 1.492 \\
\midrule
$E_{^5\mathrm{He}}$~(MeV) & -39.993 & -40.976 & -42.557 & -46.400 & -48.772 & -51.892 & -55.299 & -59.054 & -63.347&-68.276 & -73.263 \\
$E_{^4\mathrm{He}}$~(MeV) & -39.856 & -40.849 & -42.049 & -45.642 & -47.654 & -50.493 & -53.311 & -56.685 & -60.277 & -64.306 & -69.032 \\
$\kappa$~(MeV) & 0.0727& 0.0700& 0.1398& 0.1709& 0.2076& 0.2321& 0.2768& 0.3021& 0.3440& 0.3911& 0.4037 \\
\bottomrule
\end{tabular}
\end{table}

\section*{Linear formulation of the Pad\'e fitting problem}

We summarize here the explicit construction of the linear system used to determine the Pad\'e coefficients from the bound-state data.

The unknown coefficients are collected into the vector
\begin{equation}
\bm{x}=\bigl(c_1,\dots,c_N,\, d_1,\dots,d_M\bigr)^\top\in\mathbb{R}^{N+M}.
\end{equation}

For each data point, we impose
\begin{equation}
y_i\,Q_M(z_i) - P_N(z_i)=0
\ \Longrightarrow\
y_i\sum_{q=0}^{M} d_q z_i^q - \sum_{p=0}^{N} c_p z_i^p=0.
\end{equation}
Substituting $d_0=1$ and $c_0=0$ and rearranging, this relation can be written as
\begin{equation}
\underbrace{\bigl[z_i^1,\dots,z_i^N,\; -(y_i z_i^1),\dots,-(y_i z_i^M)\bigr]}_{\bm{a}_i^\top}
\begin{bmatrix}
c_1\\ \vdots\\ c_N\\ d_1\\ \vdots\\ d_M
\end{bmatrix}
= y_i,
\end{equation}

Here, $\bm{a}_i^\top$ denotes the $i$th row of the design matrix.

Collecting all $K$ data points yields the linear system
\begin{equation}
A\bm{x} \approx \bm{b},\qquad
A=\begin{bmatrix} \bm{a}_1^\top \\ \vdots \\ \bm{a}_K^\top \end{bmatrix},\quad
\bm{b}=\begin{bmatrix} y_1\\ \vdots\\ y_K\end{bmatrix}.
\end{equation}
The Pad\'e coefficients are obtained by solving the corresponding least-squares problem
\begin{equation}
\min_{\bm{x}}\ \|A\bm{x}-\bm{b}\|_2^2.
\end{equation}

\section*{SVD analysis and regularization of the Pad\'e fitting problem}

In this section, we provide technical details of the numerical stabilization procedure used to treat the ill-conditioned Pad\'e fitting problem.

The Pad\'e construction can be formulated as a linear map
$A:\mathbb{C}^{N+M}\rightarrow\mathbb{C}^{K}$ from the coefficient space to the data space.
We analyze the numerical properties of this system using singular value decomposition (SVD)~\cite{Golub1970*,Gonnet2013*}.
There exist orthonormal vectors $\{u_i\}$ in the coefficient space such that
\begin{equation}
\|A u_i\|_2=\sigma_i,\qquad 
\sigma_{\max}=\max_i \sigma_i,\qquad 
\sigma_{\min}=\min_i \sigma_i,
\end{equation}
where $\sigma_i$ denote the singular values of $A$.

Large singular values correspond to directions in which variations of the coefficients strongly affect the output and are therefore well constrained by the data.
Conversely, small singular values identify directions that are weakly constrained, along which statistical noise is strongly amplified when solving the inverse problem.
This amplification leads to large fluctuations in the fitted coefficients and manifests itself as spurious poles in the Pad\'e approximant.

To quantify the degree of ill-conditioning, we use the spectral condition number
\begin{equation}
\mathrm{cond}(A)\equiv\frac{\sigma_{\max}(A)}{\sigma_{\min}(A)}.
\end{equation}

To stabilize the fitting procedure, we apply column equilibration followed by ridge (Tikhonov) regularization~\cite{Tikhonov1963*,Hoerl1970*}.
Column equilibration rescales each column of the design matrix by its Euclidean norm, yielding an equilibrated matrix $A^{(s)}$ with comparable column magnitudes.
This procedure does not alter the column space of the matrix, but typically increases $\sigma_{\min}$ and reduces $\mathrm{cond}(A)$ by eliminating artificial scale disparities associated with different polynomial orders.

Residual near-linear dependencies are further suppressed by ridge regularization.
Specifically, we introduce a quadratic penalty term $\lambda_T\|\bm{x}\|_2^2$ and solve the regularized least-squares problem
\begin{equation}
\min_{\bm{x}}\;\bigl\|A^{(s)}\bm{x}-\bm{b}\bigr\|_2^2+\lambda_T\|\bm{x}\|_2^2,
\end{equation}
which can be written equivalently as the augmented system
\begin{equation}
\min_{\bm{x}}\left\|
\underbrace{\begin{bmatrix}
A^{(s)}\\[2pt]
\sqrt{\lambda_T}\,I
\end{bmatrix}}_{\widetilde A}
\bm{x}-
\underbrace{\begin{bmatrix}
\bm{b}\\[2pt]
0
\end{bmatrix}}_{\widetilde b}
\right\|_2^2.
\end{equation}

From the perspective of singular values, ridge regularization modifies the spectrum according to
\begin{equation}
\sigma_i \;\longmapsto\; \sqrt{\sigma_i^2+\lambda_T},
\end{equation}
thereby imposing a uniform lower bound on the singular values.
As a result, the condition number is reduced to
\begin{equation}
\mathrm{cond}(\widetilde A)
=\frac{\sqrt{\sigma_{\max}^2+\lambda_T}}{\sqrt{\sigma_{\min}^2+\lambda_T}}.
\end{equation}
Poorly constrained directions associated with small singular values are strongly damped, while directions with large singular values are only weakly affected.

\end{document}